\documentclass[runningheads]{llncs}
\usepackage{graphicx}
\usepackage{booktabs} 

\usepackage{lipsum} 
\usepackage{balance}
\usepackage{paralist}
\usepackage{filecontents}
\usepackage{etoolbox}
\usepackage{mathrsfs}
\usepackage{wrapfig}
\usepackage{bm}
\usepackage{setspace}
\usepackage{amssymb}
\usepackage{graphicx}
\usepackage[justification=centering]{caption}
\usepackage{subcaption}
\usepackage[ruled]{algorithm2e}
\usepackage{multirow}
\usepackage{amsmath, mathtools}
\usepackage{hyperref}

\let\subparagraph\paragraph
\usepackage[compact]{titlesec}

\usepackage{url}
\urldef{\mailsa}\path|{hanszeng, brutusxu, aiqy}@cs.utah.edu|
\titlespacing*{\section}{0pt}{5pt plus 1pt}{5pt minus 2pt}
\titlespacing*{\subsection}{0pt}{4pt plus 1pt}{4pt minus 2pt}

\begin{document}

\title{A Zero Attentive Relevance Matching Network for Review Modeling in Recommendation System}

\author{Hansi Zeng, Zhichao Xu, and Qingyao Ai}
\institute{School of Computing, University of Utah, Salt Lake City, UT, USA, 84112\\
	\mailsa}

\maketitle
\begin{abstract}
User and item reviews are valuable for the construction of recommender systems. 
In general, existing review-based methods for recommendation can be broadly categorized into two groups: the siamese models that build static user and item representations from their reviews respectively, and the interaction-based models that encode user and item dynamically according to the similarity or relationships of their reviews. Although the interaction-based models have more model capacity and fit human purchasing behavior better, several problematic model designs and assumptions of the existing interaction-based models lead to its suboptimal performance compared to existing siamese models.
In this paper, we identify three problems of the existing interaction-based recommendation models and propose a couple of solutions as well as a new interaction-based model to incorporate review data for rating prediction.
Our model implements a relevance matching model with regularized training losses to discover user relevant information from long item reviews, and it also adapts a zero attention strategy to dynamically balance the item-dependent and item-independent information extracted from user reviews.
Empirical experiments and case studies on Amazon Product Benchmark datasets show that our model can extract effective and interpretable user/item representations from their reviews and outperforms multiple types of state-of-the-art review-based recommendation models.

\end{abstract}

\keywords{ \noindent Review Modeling, Interaction-based Model, Relevance Matching }

\section{Introduction}
Review text is considered to be valuable for effectively learning user and item representations for recommender systems.
Previous studies show that the incorporation of reviews into the optimization of recommender systems can significantly improve the performance of rating prediction by alleviating data sparsity problems with user preferences and item properties expressed in review text~\cite{TopicIntegratedHF,TopicIntegratedRMR,DeepCoNN,narre}. 
In general, existing review based recommender systems for rating prediction can be roughly categorized into two groups: 1) The siamese models that independently encode static user and item representations from reviews and use the static representations to predict the rating~\cite{DeepCoNN,narre}; 2) The interaction-based models that dynamically learn the user and item representations based on their context~\cite{MPCN,AHN}. In particular, the interaction-based models assume that, given different target items, different user reviews might play different roles in determining the utility of the items. For example, when the target item is about an
album from the Led Zeppelin, the user review that reflects her interest on Rock $\&$ Roll music might be more useful than the rest of her reviews.

Although the interaction-based models have more model capacity and fit the human purchasing behavior better~\cite{MPCN},
several problematic model designs and assumptions lead to its lower performance than siamese models as shown in recent studies \cite{how-review}. First, most existing interaction-based models exploit co-attention mechanism \cite{qa-attentive,re2,esim,caffe} to distill textual similarity information between user and item reviews,  but such information might be diluted when there is a vast amount of text in user and item reviews. 
Second, because the number of reviews to profile each user in the training set is limited, it is common that the target item's characteristics are beyond the interest of a user expressed in her limited reviews. 
Interaction-based models that force the user representations to extract valuable information from user reviews for the target item might introduce irrelevant aspects of the user and cause serious overfitting.
Third, existing interaction-based models extract user-item relationships mostly by modeling the textual similarity between user and item reviews. 
High textual similarities between user and item reviews, however, not necessarily reflect the user's true opinion on the target item. 
For example, an item review ``the taste of cappuccino is really good" might have higher textual similarity with a user review ``I really enjoy the taste of beef pho'' than with ``I am a big coffee fan", but the latter review that reflects the user opinion on coffee could be more informative when predicting her rating on the target item. 

Based on these observations, in this paper, we propose a new interaction-based rating prediction model to mitigate the weakness of the existing interaction-based recommendation models. 
First, we implement a relevance matching model~\cite{drmm} instead of a semantic matching model~\cite{re2} to search the relevant review from the user to the target item. 
Our relevance matching model treats each user review as a query to search and extracts relevant information from all the reviews of the target item.
It is capable of discovering relevance information from a large amount of review text with thousands or more words. 
Second, to better capture the semantic relationships instead of the textual similarity between user and item reviews, we use the \textit{ground-truth} review (available in the training stage) written from the user to the target item and the corresponding item reviews as a pair of positive ``query-document'' to train our relevance matching module and plug it as the auxiliary loss in the training objective, since the \textit{ground-truth} review expresses the user true opinion to the target item. 
After the relevance matching function is well-trained, other user reviews that have high relevance matching scores to the target item would share similar characteristics to the \textit{ground-truth} review and also reflect the user true interest to the target. 
Last but not least, when there is not relevant review from the user to the target item, we exploit a zero-attention network \cite{zero-attention} to avoid using irrelevant reviews to build user representations.
Our zero-attention network not only builds dynamic user representations when there are high informative reviews from the user to the target item, but also allows the model to degenerate to a siamese model with static user representations when all user reviews are not relevant to the target item. 
Specifically, separated from the interaction module, we build static user and item embeddings using a multi-layer convolutional self-attention network to extract information hierarchically from words, sentences and reviews.
We then construct the final user representations using both the dynamic user representations extracted by the interaction module and the static user embeddings created by the self-attention network.
When there is no user review relevant to the target item, the dynamic user representation created by the interaction module with zero attention networks would be downgraded to a zero vector and the final rating prediction of the user-item pair would purely depend on the static user and item embeddings.
Empirical experiments and case studies on four datasets from Amazon Product Benchmark show that that our proposed model the Zero Attentive Relevance Matching Network (ZARM) can extract effective and interpretable user/item representations from review data and outperforms multiple types of state-of-the-art review-based recommendation models.

\section{Related Works}

\textbf{\underline{Review Based Recommendation}.}
Using review text to enhance user and item representations for recommender system has been widely studied in recent years \cite{TopicIntegratedHF,TopicIntegratedRMR,Zhang2014ExplicitFM,He2015TriRankRE,Tan2016RatingBoostedLT,Ren2017SocialCV,Zhang2016CollaborativeME,Diao2014JointlyMA}. Many works are focus on topic modeling from review text for users and items. For example, the HFT\cite{TopicIntegratedHF} uses LDA-like topic modeling to learn user and item parameters from reviews. The likelihood from the topic distribution is used as a regularization for rating prediction by matrxi factorization(MF). The RMR\cite{TopicIntegratedRMR} uses the same LDA-like model on item reviews but fit the ratings using Guassian mixtures but not MF-like models. 
Recently, with the advance of deep learning, many recommendation models start to combine neural network with review data to learn user and item representations, including DeepCoNN\cite{DeepCoNN}, TransNets\cite{Catherine2017TransNetsLT} D-Att\cite{cnndlga}, NARRE\cite{narre},  HUITA\cite{HUITA}, HANN\cite{HANN}, MPCN\cite{MPCN}, AHN\cite{AHN}, HSACN\cite{zeng2020hierarchical}. 
Despite their differences, existing work using deep learning models for review modeling can be roughly divided into two styles -- siamese networks and interaction-based networks. For example, DeepCoNN\cite{DeepCoNN} uses two convolution neural networks to learn user and item representations from reviews statically; NARRE\cite{narre} extends CNN with an attention network over review-level to select reviews with more informativeness. In addition, the MPCN\cite{MPCN}, a interaction-based network, uses co-attention mechanism to select the most informative and matching reviews from the user and item respectively, then another attention mechanism is applied to learn the fixed dimensional representation, by modeling the word-level interaction on the matched reviews. However, both of these two styles have their own weaknesses. The siamese models lack the dynamic target-dependent modeling and neglect the interaction between the user and target. But the interaction-based models forcely require the dynamic matching between each user and item, neglecting the fact that not every user exists the informative review to the target. Even the informative review exists, the matching information might be diluted considering thousands of words within tens of review for profiling the user and item.

\noindent \underline{\textbf{Interaction Based Text Matching}.}
The review based dynamic user-item modeling is closely related to query-document representation learning in the QA \cite{qa-attentive} task or premise-hypothesis encoding in the NLI \cite{re2,esim,caffe} task that exploit the co-attention mechanism. The co-attention mechanism computes the pairwise similarity between two sequences, builds the pair-wise attention weights, and integrates them with other feautres of the sequences for effective text semantic matching learning.
Besides the text semantic matching in the NLP tasks, several works on the IR tasks\cite{drmm,knrm,guo2020deep,bi2020trans} also utilize the interaction-based approaches for text relevance matching learning. For example, DRMM \cite{drmm} proposes a pooling pyramid technique that converts the pair-wise similarity matrix into the histogram, and use it as feature for final text matching prediction. Based on the DRMM, K-NRM \cite{knrm} introduces the kernel-based differential pooling technique that can learn the matching signal in different level.
Recent work\cite{Rao2019BridgingTG} further investigate using the semantic matching and relevance matching together or alone in the NLP and IR tasks. It finds that using relevance matching alone performs reasonable well in many NLP tasks but the semantic matching is not effective for IR tasks.

\noindent \underline{\textbf{Rethinking the Progress of Deep Recommender System}.}
While we have witnessed the rapid advancements of deep learning methodology and its applications on the field of recommender systems, there are worries about the progress we made. Dacrema et al.\cite{AreWeReallyMakingProgress} investigated the performance of several recent algorithms proposed in top conferences and found most of them can not compete with traditional methods, like Matrix Factorization and its derivative models \cite{koren2009matrix,neurMF,xu2020commerce}, BPR \cite{rendle2012bpr} or Item-KNN. Furthermore, Sachdeva et al.\cite{how-review} focused on the usefulness of reviews. He examined several review-based recommendation algorithms and found that 
applying complex structures to extract semantic information in reviews, not necessarily improve the system's performance. Our goal is to try to tackle these existing problems in this field and propose an interpretable method to effectively utilize the review information.
\section{Proposed Method}
\setlength{\belowdisplayskip}{0pt} \setlength{\belowdisplayshortskip}{0pt}
\setlength{\abovedisplayskip}{0pt} \setlength{\abovedisplayshortskip}{0pt}
\subsection{Overview}
\begin{figure}[t]
   \centering
    \includegraphics[width=0.85\textwidth, height=0.60\textwidth]{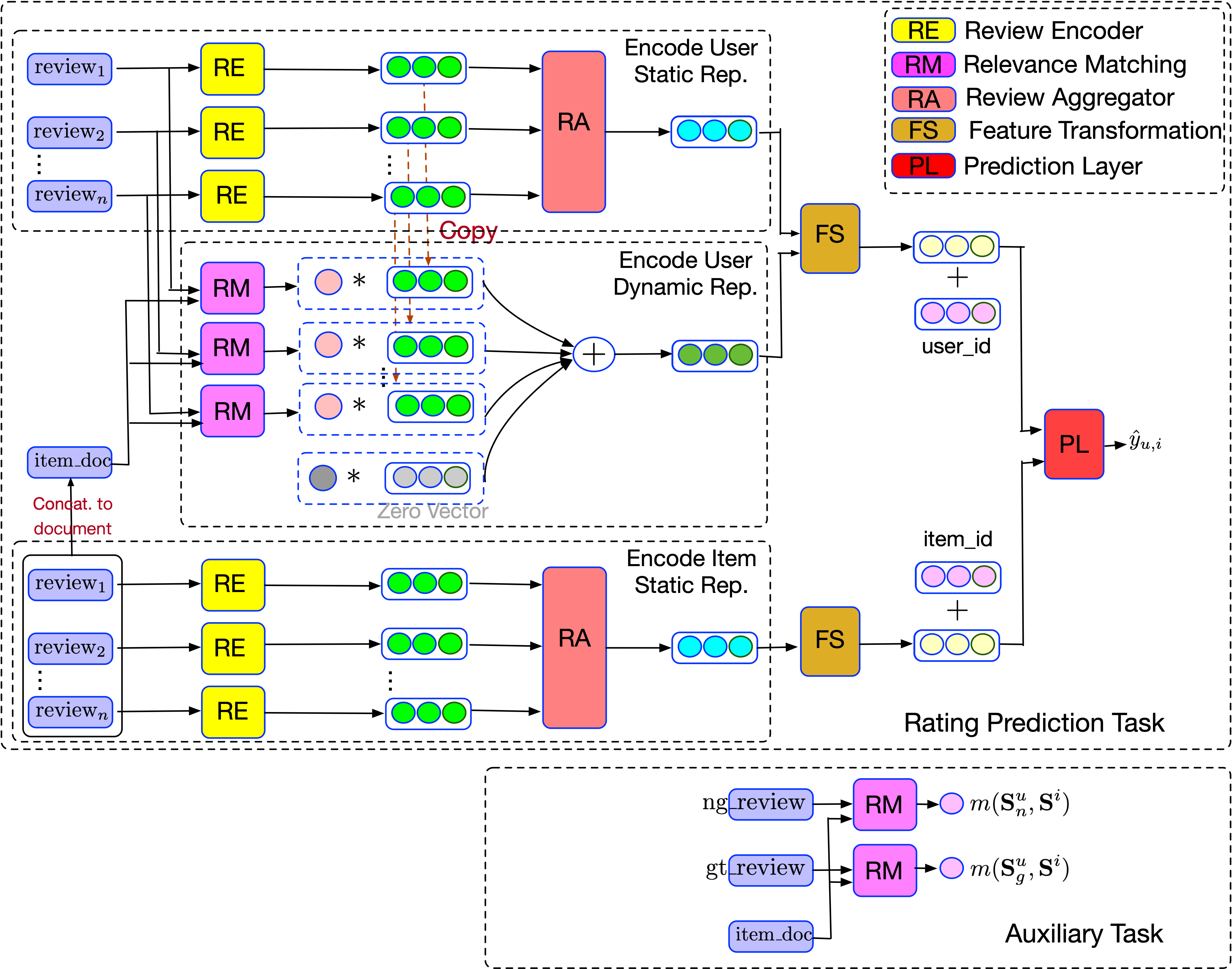}
    \caption{Overview of our model structure}
   \label{fig:arch}
   \vspace{-26pt}
\end{figure}

The goal of the proposed model is to predict the rating from the user to the target item based on their review text. The architecture of our model is shown in Figure. \ref{fig:arch}. Our model contains two parallel encoders that use multi-layer convolution self-attention network to hierarchically encode user and item static representations from their reviews respectively. Besides, the model has a interaction module that encodes the user dynamic representation according to her current interacted item where we first compute the relevant level of each user review to the target item by the relevance matching function. Then the zero-attention network is applied to allow the dynamic user representation degrade to a zero vector when there is no user review relevant to the target, in which case the final user representation would purely depend on the static user representation. 
The encoded user static and dynamic representations will be concatenated and be taken as the input to the feature transformation layer to encode the final user representation. On the rightmost of the model, the prediction layer is added to let the learned user and item final representations interact with each other and compute the final rating prediction. In the training stage, the auxiliary loss is plugged to guide the training of relevance matching function. In the following sections, we will introduce the static user/item encoder (section \ref{sec:static enc}), dynamic user encoder composed of the relevance matching function and zero-attention network (section \ref{sec:dyn enc}), prediction layer(section \ref{sec:pred}), and the training objective (section \ref{sec: training}) in details.
\subsection{Static User/Item Encoder} \label{sec:static enc}
Since the static user and item encoder only differ in their inputs, we introduce the process of encoding user static representation in the following in details. And the same process is applied to static item encoder in the similar way. Assume the input of the user encoder is $\{r^u_1, \ldots r^u_N \}$, where $N$ is the number of reviews written by the user. We learn each review representation hierarchically from word-level to sentence-level. More specifically, a user review $r^u = \{s_1, \ldots s_T\}$ consists of $T$ sentences, and each sentence $s_i$ is composed of a sequence of $L$ words $ \{w^i_1, \ldots w^i_L \}$. To learn the sentence representation $\bm{s_i}$, we apply the word-level self-attentive convolution network to encode the contextual representation of each word in the sentence and use the attention network to aggregate the learned contextual embeddings in to a single vector. Mathematically, we first apply the word embedding layer to map each word $w^i_j$ into a vector $\bm{w}^i_j \in \mathbb{R}^{d_w}$ to form a sequence of word embeddings $\bm{W}^i \in \mathbb{R}^{d_w \times L}$, then we apply the word-level convolution neural network to learn the local semantic representation of each words: 
\begingroup\makeatletter\def\f@size{9}\check@mathfonts
\begin{align}
    \bm{Q}^i_w = \text{CNN}_w^Q (\bm{W}^i), \ \bm{K}^i_w = \text{CNN}_w^K (\bm{W}^i) ,\ \bm{V}^i_w = \text{CNN}_w^V (\bm{W}^i)
\end{align}
\endgroup

where $\bm{Q}^i_w ,\ \bm{K}^i_w ,\ \bm{V}^i_w \in \mathbb{R}^{d_w \times L}$. To enrich each word semantic representation and capture long-range dependencies between words, we apply the multihead-self-attention network~\cite{attention} on top of the learned word local representation from $\text{CNN}(\cdot)$. Finally, a $1$ layer feed-forward network is sequentially plugged to learn more flexible representations:  
\begingroup\makeatletter\def\f@size{9}\check@mathfonts
\begin{align}
    \bm{Z}_w^i = \text{FFN}_w \bigg( \text{Multihead-Self-Attention}_w(\bm{Q}^i_w, \bm{K}^i_w, \bm{V}^i_w) \bigg)
\end{align}
\endgroup
where $\bm{Z}_w^i \in \mathbb{R}^{d_s \times L}$. Then we use addictive-attention network \cite{Yang2016HierarchicalAN} to aggregate the contextual representations into a single vector $\bm{s_i} \in \mathbb{R}^{d_s}$ for sentence modeling:
\begingroup\makeatletter\def\f@size{9}\check@mathfonts
\begin{align}
   \bm{s_i} = \text{Addictive-Attention}_w (\bm{Z}^i_w)
\end{align}
\endgroup
We apply the same procedure on each sentence of the review $r^u$ to form a sequence of sentence representations $\bm{S} = [\bm{s}_1, \ldots \bm{s}_T] \in \mathbb{R}^{d_s \times T}$. Then we take the sentence sequences as an input to sentence-level self-attentive convolution network with addictive-attention network to form the review representation $\bm{r}^u \in \mathbb{R}^{d_r}$:
\begingroup\makeatletter\def\f@size{9}\check@mathfonts
\begin{align}
    \bm{Z}_s &= \text{Multihead-Self-Attention}_s \bigg( \text{CNN}^Q_s(\bm{S}), \text{CNN}^K_s(\bm{S}), \text{CNN}^V_s(\bm{S}) \bigg) \\
    \bm{r}^u &= \text{Addictive-Attention}_s (\bm{Z}_s)
\end{align}
\endgroup
We apply the same hierarchical network on each review written by the user, then form a sequence of review representations $\bm{R} = [\bm{r}^u_1, \ldots \bm{r}^u_N] \in \mathbb{R}^{d_r \times N}$. Finally, we apply the user-level addictive-attention network to aggregate the information of these reviews and form a single vector $\bm{u}^{static} \in \mathbb{R}^{d_r}$ to form the user static representation:
\begingroup\makeatletter\def\f@size{9}\check@mathfonts
\begin{align} \label{eq:rev_aggr}
    \bm{u}^{static} &= \text{Addictive-Attention}_u (\bm{R}^u)
\end{align}
\endgroup
The item static representation $\bm{i}^{static}$ can be obtained using a similar procedure.
\vspace{-12pt}
\subsection{Dynamic User Encoder}\label{sec:dyn enc}
\vspace{-3pt}
To learn the dynamic user representation, we first compute the relevant scores of the reviews of the user to the target item using relevance matching function. In other words, given $N$ reviews written by the user $\{ r^u_1, \ldots r^u_N \}$, we want to compute their corresponding relevance scores $\{ \alpha_1, \ldots \alpha_N \}$ to the target. The detailed introduction of the relevance matching function is in the following.
\\
\underline{\textbf{Relevance Matching Function}}:
The input of the function is a query-document pair where we treat the user review as a query and the target item reviews as a document, and we denote the function as $m(\cdot, \cdot)$. Formally, each user review $r^u_k$ can be alternatively represented as a sequence of word embeddings $[ \bm{e}_1^u, \ldots, \bm{e}_M^u ] := \bm{S}_u^k$, and the item document is a concatenation of a sequence of word embeddings of its each review  $[ \bm{e}_1^i, \ldots, \bm{e}_M^i, \ldots, \bm{e}_{(N-1)M +1}^i, \ldots, \bm{e}_{NM}^i] := \bm{S}_i$, where $\bm{e}^u_k, \ \bm{e}^i_k \in \mathbb{R}^{d_w}$, $\bm{S}^k_u \in \mathbb{R}^{d_w \times M}$,  $\bm{S}^i \in \mathbb{R}^{d_w \times MN}$, $d_w$ is the dimension of the word embedding, $M$ is the review length, and $N$ is the number of review from the target item. To get the relevant matching score from the $k$-th user review to its target item, we first compute the word similarity matrix $S$:
\begingroup\makeatletter\def\f@size{9}\check@mathfonts
\begin{align}
    \bm{M} = {\bm{S}^k_u}^T \bm{S}_i \in \mathbb{R}^{M \times MN} 
\end{align}
\endgroup
where $\mathbf{M}_{i,j}$ can be considered as cosine similarity score (we normalize it into cosine space) by matching $i$-th word of user review with $j$-th word of item document. We apply mean pooling and max pooling on every row of similarity matrix to obtain discriminate features:
\begingroup\makeatletter\def\f@size{9}\check@mathfonts
\begin{align}
    mean(\mathbf{M}) = \begin{bmatrix} mean(\mathbf{M_{1:}}) \\  \ldots  \\ mean(\mathbf{M_{n:}}) \end{bmatrix} \in \mathbb{R}^M \ , 
    max(\mathbf{M}) = \begin{bmatrix} max(\mathbf{M}_{1:}) \\  \ldots  \\ max(\mathbf{M}_{n:}) \end{bmatrix} \in \mathbb{R}^M
\end{align}
\endgroup
Also, we consider the relative important score for each word in the user review $\bm{S}^k_u$ by applying a function $imp(\cdot)$:
\begingroup\makeatletter\def\f@size{9}\check@mathfonts
\begin{align}
    imp({\mathbf{S}_u^k}) = \begin{bmatrix} 
              imp({\mathbf{e}_l^u}) \\
              \ldots \\
              imp({\mathbf{e}_l^u})
    \end{bmatrix} \in \mathbb{R}^M \ \ \text{where}, \ \
    imp(\bm{e}_j^u) = \frac{\exp({\bm{w}_{p}^T \bm{e}_k^u)}}{\sum_{o=1}^n \exp({\bm{w}_{p}^T} \bm{e}_o^u)}
\end{align}
\endgroup
where $\bm{w}_{p} \in \mathbb{R}^{d_w}$, then the input feature for scoring function parameterized by a $2$ layer feed-forward neural network is:
\begingroup\makeatletter\def\f@size{9}\check@mathfonts
\begin{align}
    \bm{I}^{rel} =  \begin{bmatrix} 
        imp(\bm{S}^k_u) \odot mean(\mathbf{M}) \\
        imp(\bm{S}^k_u) \odot  max(\mathbf{M})
    \end{bmatrix} \in \mathbb{R}^{2M}
\end{align}
\endgroup
Hence the relevant score between the $k$-th user review $\bm{S}^u_k$ and item document $\bm{S}^i$ is:
\begingroup\makeatletter\def\f@size{9}\check@mathfonts
\begin{align}
    m(\bm{S}^u_k, \bm{S}^i) = \text{FFN} \big( \text{FFN}(\bm{I}^{rel}) \big) = \alpha_k \in [-\infty, \infty] \label{eq:matching} 
\end{align}
\endgroup
When there is no user review relevant to the target item, we can expect each relevant score $\alpha_k \ll 0$. However, if we naively normalized the relevant scores, and use them as weights to measure the importance of each user review, the final dynamic user representation we get by weighted sum of the user review representations would be a non-zero vector. It is due to the fact that after the normalized process, every relevant score will be assigned as a probability measure, and summation of these probabilities being $1$ makes the situation that every normalized relevant weight $a_k \approx 0$ become impossible, hence the weighted sum of the user reviews cannot be a zero vector. 
For example, suppose that the relevant scores of all user reviews are $\alpha_k = -100$, where $k=1, \ldots N$, then the normalized relevant score would be $\alpha_k = \frac{1}{N}$. 
The dynamic item representation will become $\bm{u}^{dynamic} = \displaystyle\sum_{k=1}^N \frac{1}{N} \bm{r}^u_k$, which is not a zero vector even when all user reviews are not relevant to the target item. To resolve the problem, we use the zero-attention network motivated by \cite{zero-attention}.\\
\noindent\underline{\textbf{Zero-Attention Network}}: we introduce a zero score $\alpha_0 = 0$, and re-normalize the relevant scores by taking the zero score in to account.
Formally, $\hat{\alpha}_k = \frac{\exp{(0)} + \exp(\alpha_k)}{\exp{(0)} + \exp{(\alpha_1)} + \ldots \exp{(\alpha_m)}} = \frac{1 + \exp(\alpha_k)}{1 + \exp{(\alpha_1)} + \ldots \exp{(\alpha_m)}}, \ k=1, \ldots N $, and $\hat{\alpha}_0 = \frac{1}{1 + \exp{(\alpha_1)} + \ldots \exp{(\alpha_m)}}$, then the user dynamic representation is,
\begingroup\makeatletter\def\f@size{9}\check@mathfonts
\begin{align}
    \mathbf{u}^{dynamic} &= \displaystyle\sum_{k=1}^N \hat{\alpha}_k \mathbf{r}^u_k + \hat{\alpha_0} \vec{\mathbf{0}} 
\end{align}
\endgroup
Intuitively, when $\alpha_k \ll 0$, the normalized score $\hat{\alpha}_k \approx 0$ for all $k=1, \ldots, N$, and the $\bm{u}^{dynamic} \approx \vec{\mathbf{0}}$, which is close to a zero vector. In the other hand, if there exist a large relevant score, for example $\alpha_k = 10$ for a certain $k$, the effect of $\alpha_0 = 0$ will be very low, and the normalized score will be $\hat{\alpha}_k \approx 1$, and $\bm{u}^{dynamic} \approx \bm{r}^u_k$
\subsection{Prediction Layer} \label{sec:pred}
This layer combine the static and dynamic user representations to form a final user representation learned from reviews. Also, it learns a final item static representation from reviews by $1$-layer feed-forward neural network:
\begingroup\makeatletter\def\f@size{9}\check@mathfonts
\begin{align}
    \bm{u}^r &= \text{Relu} \Big( \begin{bmatrix} 
    \mathbf{W}^{static}_u, & \mathbf{W}^{dynamic}_u
     \end{bmatrix}\begin{bmatrix}
        \bm{u}^{static} \label{eq:u} \\
        \bm{u}^{dynamic} 
        \end{bmatrix}  + \mathbf{b}_u \Big) \\ 
    \bm{i}^r &= \text{Relu} \Big(  \bm{W}^{static}_i \bm{i}^{static} + \mathbf{b}_i \Big) \label{eq:i}
\end{align}
\endgroup
where $\bm{W}^{static}_u, \ \bm{W}^{static}_u ,\ \bm{W}^{dynamic}_i, \ \bm{W}^{static}_i \in \mathbb{R}^{d_h \times d_r}$, $\bm{b}_u, \ \bm{b}_i \in \mathbb{R}^{d_h}$. Finally, we combine the user and item id embeddings $\bm{u}^{id}$, $\bm{i}^{id} \in \mathbb{R}^{d_h}$, with user and item embeddings learned from reviews $\bm{u}^{r}$, $\bm{i}^{r}$, to form their final representations, which are  $\bm{u} = \bm{u}^r + \bm{u}^{id} ,\ \bm{i} = \bm{i}^r + \bm{i}^{id}$. We take the user and item embeddings as input to get the final rating prediction:
\begingroup\makeatletter\def\f@size{9}\check@mathfonts
\begin{align}
    \hat{y}_{u,i} = \bm{w}_f^T (\bm{u} \odot \bm{i}) + b_u + b_i + b_g \label{eq:rating} 
\end{align}
\endgroup
where $\bm{w}_f \in \mathbb{R}^{d_h}, \ b_u, \ b_i, \ b_g \in \mathbb{R}$ 

\subsection{Training Objective}\label{sec: training}
Besides a regression loss for the rating prediction, an auxiliary loss is utilized for better training the relevance matching function. Specifically, we assumed there is a user-item pair $(u, i)$ with the \textit{ground-truth} rating $y_{u,i}$ in the training stage, and the \textit{ground-truth} review $r_{u,i}^g$ written from the user to the target item is treated as a ``positive query" to the target item. Also we randomly sample a review from the different user different item as a "negative query" to the target item which is $r_{u,i}^n$. The corresponding word sequence representation of the \textit{ground-truth} review, negative review and target item document is $\bm{S}^u_g, \ \bm{S}^u_n, \ \bm{S}^i$. Ideally, a good relevance matching function $m(\cdot, \cdot)$ can distinguish the positive query-document pair from the negative one, in other words, we wish $ m (\bm{S}^u_g,  \bm{S}^i) > m(\bm{S}^u_n,\bm{S}^i)$. In the same time, we want to minimize the regression loss between ground-truth rating $y_{u,i}$ and predicted rating $\hat{y}_{u,i}$ computed from Equation \ref{eq:rating}. To achieve the above two goals, we write the objective function as followed,
\begingroup\makeatletter\def\f@size{9}\check@mathfonts
\begin{align*}
     loss = \displaystyle\sum_{\{ (u,i)\} \in \mathcal{S}} \underbrace{\bigg( y_{u,i} - \hat{y}_{u,i} \bigg)^2}_{\text{regression loss}} - \underbrace{\bigg( \log\big( m (\bm{S}^u_g,  \bm{S}^i)\big) +  \log\big(1 - m(\bm{S}^u_n,\bm{S}^i) \big)\bigg)}_{\text{auxiliary loss}} \label{eq:loss}
\end{align*}
\endgroup
\setlength{\textfloatsep}{5pt plus 1.0pt minus 2.0pt}
\setlength{\intextsep}{5.0pt plus 2.0pt minus 2.0pt}
\vspace{-1.1em}
\section{Experimental Setup}
\vspace{-.4em}
\underline{\textbf{Datasets and Evluation Metrics}.}
We conduct our experiment on four different categories of 5-core Amazon product review datasets \cite{amazon-dataset}. The statistics of these four categories are shown in the first and second columns of the Table \ref{tab:perfom_comp}. 
For each dataset, we randomly split user-item pairs into training, validation, and testing sets with ratio 8:1:1. We use NLTK\cite{NLTK} to tokenize sentences and words of reviews. We let the number of reviews 
be the same for profiling user and item where the number of reviews is set to cover $90\%$ 
of users  for the balance of efficiency and performance. We adopt Mean Square Error (MSE) as the main metric to evaluate the performance of our model. The source code can be found here  \footnote{\small https://github.com/HansiZeng/ZARM}.

\noindent \underline{\textbf{Compared Methods}.}
To evaluate the performance of our method, we compare it to several state-of-the-art baseline models: (1) \textbf{MF} \cite{mf}: a basic but well-known CF model that predict the rating using inner product between user, item hidden representations plus user, item and global bias; 
(2) \textbf{NeurMF} \cite{neurMF}: the CF based model combines linearity of GMF and  non-linearity of MLPs for modeling user and item latent representations; 
(3) \textbf{HFT} \cite{TopicIntegratedHF}: the topic modeling based model combines the ratings with reviews via LDA; (4) \textbf{DeepCoNN} \cite{DeepCoNN}: the CNN based model uses two convolution neural network to learn user and item representation; 
(5) \textbf{NARRE} \cite{narre}: the CNN based model modifies the DeepCoNN by using the attention network over review-level to select reviews with more informativeness.
(6) \textbf{MPCN} \cite{MPCN}: the model that selects informative reviews from user and item by review-level pointers using the co-attention technique, and selects informative word-level representations for the rating prediction by applying word-level pointers over the selected reviews; 
(7) \textbf{AHN} \cite{AHN}: a dynamic model using co-attention mechansim but treats user and item asymmetrically; 
(8) \textbf{ZARM-static}: the variant of the ZARM that only user static representations; And (9) \textbf{ZARM-dynamic}: the variant of the ZARM that only uses user dynamic representations.   

\noindent \underline{\textbf{Parameter Settings}.}
We use 300-dimensional pretrained word embeddings from Google News \cite{googlenews}, and employ the Adam\cite{Kingma2015AdamAM} for optimization with an inital learning rate $0.001$. We set the dimension of sentence hidden vector and review hidden vector as $100$, and the latent dimension of the prediction layer as $32$. Also, the convolution kernel size is $1$ or $3$ based on the performance in each dataset, and number of head for each self-attention layer is $2$. We apply dropout after the word embedding layer, after each feed forward layer in sequence encoding modules, and before the prediction layer with rate $[0.2, 0.3, 0.5]$. 
The hidden dimension of the two layer neural network in the Relevance Matching Module is set to $16$. 
The hyper-parameters of baselines are set following the settings of their original papers.

\setlength{\textfloatsep}{5pt plus 1.0pt minus 2.0pt}
\setlength{\intextsep}{5.0pt plus 2.0pt minus 2.0pt}

\section{Results and Analysis}
The MSE results of compared models are shown in Table \ref{tab:perfom_comp}.
Based on the results, we can make several observations. 
Firstly, the siamese models outperform the interaction-based models significantly.
As discussed previously, due to the fact that not every user exists informative review to the target, interaction-based models that force to extract informative reviews from user data will suffer from heavily over-fitting. 
Among siamese networks, we observe that the ZARM-static outperforms the other siameses models. 
This demonstrates that ZARM-static can capture the review hierarchical structure and use attention neural network to select the important information in each level. 
Among interaction-based models, ZARM-dynamic outperforms the other baselines such as MPCN and AHN. 
This demonstrates the effectiveness of the relevance matching component in discovering relevant information from vast review text and the utility of the auxiliary training loss that makes the found relevant review more aligned with the \textit{ground-truth} that reflect the user true opinion on the target item. 
Finally, our model (i.e., ZARM) shows consistently improvement over siamese and interaction-based models across all datasets.
Our model uses the zero-attention network that can build dynamic user representations from reviews when there are high informative reviews and can easily degrade to user static representations when there is not. 
This strategy combines the advantages of both the siamese and interaction-based models.

\begin{table}[t]
	\begin{center}
		\caption{\small Experiment results on benchmark datasets. $\dag$ and $\ddag$ represents the best performance among siamese and interaction-based models, respectively. The bold value is the best performance among all models in each dataset.}
		\label{tab:perfom_comp}
		\resizebox{\textwidth}{!}{%
			
			\begin{tabular}{c|c|c|c|c|c}
				\hline
				\multicolumn{2}{c|}{\begin{tabular}[c]{@{}c@{}}Dataset \\ \#Reviews / \#Users / \#Items\end{tabular}} & \begin{tabular}[c]{@{}c@{}}Toys \& Games \\ 167k / 19k / 12k\end{tabular} & \begin{tabular}[c]{@{}c@{}}Video Games \\ 232k / 24k / 11k\end{tabular} & \begin{tabular}[c]{@{}c@{}}Kindle Store \\ 983k / 68k / 62k\end{tabular} & \begin{tabular}[c]{@{}c@{}}Office Products \\ 53k / 2k / 4k\end{tabular} \\ \hline
				\multirow{2}{*}{Non-text-based} & MF & 0.8010 & 1.0979 & 0.6231 & 0.6954 \\ 
				\cline{2-2}
				& NeuMF & 0.8012 & 1.0931 & 0.6255 & 0.6941 \\ 
				\hline
				\multirow{4}{*}{Siamese} 
				& HFT & $0.7947^{\dag}$ & 1.0837 & 0.6172 & 0.6881 \\ 
				\cline{2-2}
				& DeepCoNN & 0.8273 & 1.1241 & 0.6437 & 0.7102 \\ 
				\cline{2-2}
				& NARRE & 0.7982 & 1.0881 & 0.6199 & 0.6794 \\ 
				\cline{2-2}
				& ZARM-static & 0.7952 & $1.0774^{\dag}$ & $0.6159^{\dag}$ & $0.6757^{\dag}$ \\ 
				\hline
				\multirow{3}{*}{Interaction-based}
				& MPCN & 0.8199 & 1.1062 & 0.6337 & 0.7101 \\ 
				\cline{2-2}
				& AHN & 0.8233 & 1.1137 & 0.6341 & 0.7341\\ 
				\cline{2-2}
				& ZARM-dynamic & $0.8054^{\ddag}$ & $1.1054^{\ddag}$ & $0.6279^{\ddag}$ & $0.7024^{\ddag}$\\ 
				\hline
				Hybrid & ZARM & \textbf{0.7881} & \textbf{1.0632} & \textbf{0.6083} & \textbf{0.6695}\\ 
				\hline
			\end{tabular}%
		}
	\end{center}
\end{table}

\subsection{Ablation Studies}
We conduct the ablation study on the validation sets of the the four benchmark datasets. We report the performance of $6$ variant models from the defualt model setting: 
(1) we change the static review aggregator in Equation \ref{eq:rev_aggr} to max pooling; (2) we encode each review using average embedding of words; (3) We use the  relative position representations \cite{relative-attention} to encode the relative position between entities in the self-attention network, now we remove the position encoding vectors to conduct ablation study; (4) we remove the user and item bias in Equation \ref{eq:rating}; (5) we remove the auxiliary loss in the training objective; And (6) we make our user and item representations symmetrically by adding the dynamic item encoding to represent the item.

\begin{table}[t]
	\vspace{-30pt}
	\caption{\small Ablation Study (validation MSE) on four datasets}
	\vspace{-10pt}
	\centering
	\begin{center}
		\resizebox{\columnwidth}{!}{
			\begin{tabular}{ c || c c c c }
				\toprule 
				\hline
				Architecture & Toys-and-Games & Video-Games  & Kindle-Store & Office-Product \\
				\hline 
				Default & 0.7897 & \textbf{1.0611} & \textbf{0.5961} & \textbf{0.6731} \\ \hline
				(1) max pooling & 0.7922 & 1.0645 & 0.6075 & 0.6795\\ \hline 
				(2) avg embedding & \textbf{0.7854} & 1.0641 &0.6043 & 0.6742 \\ \hline
				(3) Remove pos. vec. & 0.7913 & 1.0654 & 0.5985 & 0.6755 \\ \hline 
				(4) Remove u/i bias & 0.8021 & 1.0713 & 0.6022 & 0.6761 \\ \hline 
				(5) Remove aux. loss  & 0.8147 & 1.0944 & 0.6189 & 0.6893 \\ \hline
				(6) Add item dyn. & 0.7938 & 1.0695 & 0.6053 & 0.6800 \\ \hline
			\end{tabular}
		}
		\label{tab:ablation}
	\end{center}
\end{table}

As shown in Table \ref{tab:ablation}, the performance of ZARM would drop when we use max pooling in the aggregator, remove the position encoding vectors in self-attention network, or remove the u/i bias.
Using average word embeddings for review embeddings achieves suboptimal performance on most datasets, but it also outperforms the default ZARM on Toys-and-Games, which indicates that such simple aggregators may have some value on specific data types.
Interestingly, in our experiments, the variant architecture that encodes item using its dynamic and static representation underperforms the default ZARM which only use the item static encoding. 
This indicates that building interaction-based representations on the item side may not as profitable as they are on the user side, or the current interaction module is not suitable for the construction of dynamic item representations.

\subsection{Behavior of the Dynamic Interaction Matching}
\begin{wrapfigure}[12]{r}{0.5\columnwidth}
    \vspace{-5pt}
		\includegraphics[width=2.2in, ]{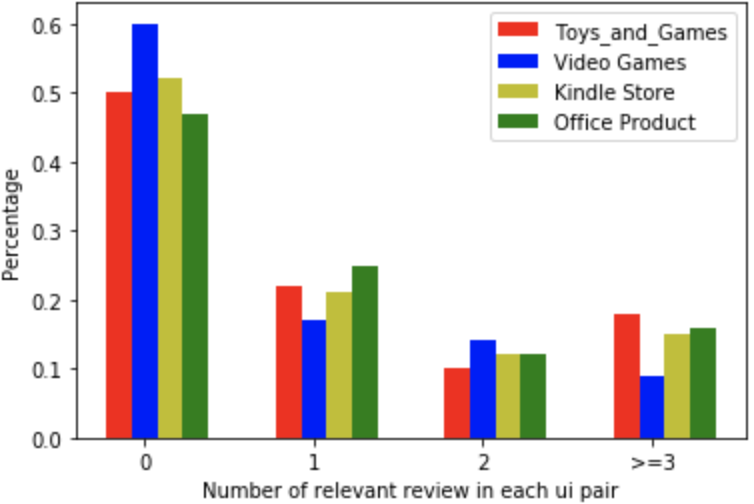}
		\vspace{-10pt}
		\caption{The distribution of user-item pairs with different numbers of relevant review ($\alpha_k>0$).}
		\label{fig:hist_com}
\end{wrapfigure}
We conduct several experiments on investigating the behaviors of the dynamic interaction matching
Firstly, we investigate the number of relevant reviews ($\alpha_k>0$ in Eq.~(\ref{eq:matching})) each user have to the target item as shown in Figure \ref{fig:hist_com}. Although the number of relevant review from the user to the target is dataset dependent, there are around $50\%$ of user-item pairs do not have the relevant review in each dataset where the type of pairs in Video Games dataset account for $60\%$ the most, and in Office Product account for $48\%$ the least. On the other hand, some users have more than one relevant reviews to their target items. For example, in the Toys $\&$ Games dataset, $15\%$ of the user-item pairs have relevant review more than 3. Such observation implies that some users have consistent interests and tend to buy items with similar characteristics, which lead to their target item matched to her multiple history items in high possibility.

\begin{figure}[t]
    \setlength{\abovecaptionskip}{3pt}
    \setlength{\belowcaptionskip}{3pt}
		\includegraphics[width=\columnwidth, height=0.3\columnwidth]{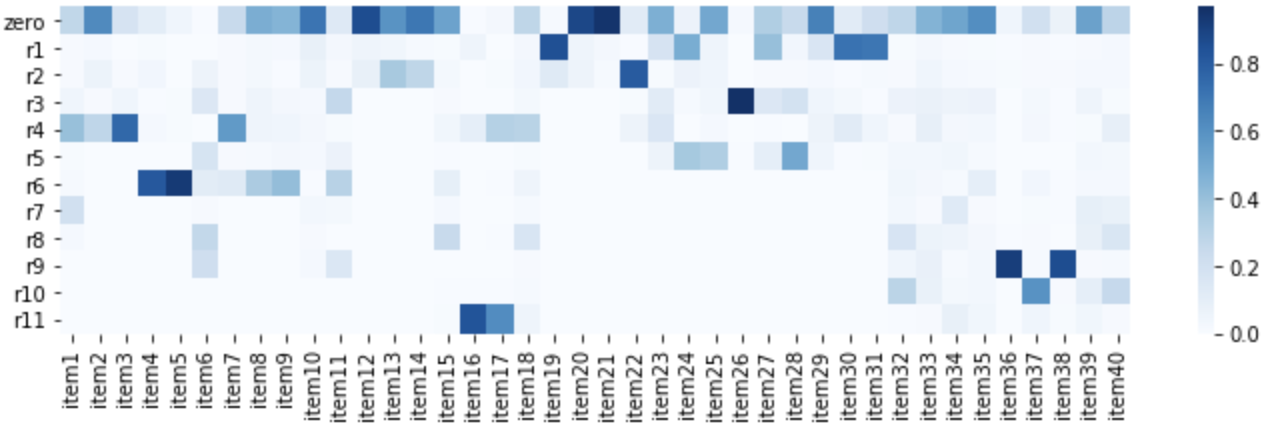}
		\vspace{-5pt}
	\caption{The distribution of relevance matching scores of each user review given a user-item pair.
	}
	\label{fig:ui_dist}
\end{figure}

To further analyze the interaction module in our model, we randomly sample $40$ users and their corresponding target items in the validation set for case studies. For each user we visualize the zero score and the relevant score of each review to the target item (from r1 to r11) as shown in Figure \ref{fig:ui_dist}. We observe that there are roughly half of the user-item pairs having large zero score which is larger than $0.5$. On the other hand, there are some users containing reviews with high relevant scores like the pair (user5, item5), (user36, item36) with review id r6, r9. We then take a closer look into the two high relevant review r6, r9 and their corresponding target item documents as shown in Table \ref{tab:rel_review}. We observe that the high relevant reviews and their target item documents share multiple similar keywords, and these keywords are highly informative that can describe the item characteristics to a large extent. For example, the keyword "Gyro Hercules" in the r6 and "Gyro Hercules helicopter" in its corresponding target item document have high textual similarity and describe the general characteristics of the two items that that r6 and target item belong to. Moreover, The user true opinion on the target item can be reflected in the high relevant reviews from the user to target. For example, the first target item (item5) which has the advantage of "keep on going and not falls down" meets the user interest that is shown in r6 that mentions she likes a helicopter that is "truly withstand a hard fall". And the second target item (item 36) which is suitable for kid Christmas gift conforms to the user interest reflected in r9 in which she mention that she needs a Christmas gift for her 3-year-old granddaughter.


\begin{table}[t]
    \setlength{\abovecaptionskip}{3pt}
    \setlength{\belowcaptionskip}{3pt}
    \centering
    \caption{Examples of high relevant reviews r6, r9 and their corresponding target item documents. The first column is their complete user review, and the second column are sampled text selected from the item documents.}
     \begin{tabular}{ |  p{0.5\columnwidth} | p{0.5\columnwidth} | }
	\hline
     User Review & Target Item Document \\ \hline 
          \scriptsize I have bought other remote \textbf{control helicopters} only to take them outside and have a little breeze of wind \textbf{knock it down and break}. With the \textbf{Gyro Hercules} it can \textbf{truly withstand a hard fall} so you can fly it nearly anywhere. & \scriptsize My kids demolish other helicopters / \textbf{keeps on going and not falls down} / helicopter / \textbf{Gyro Hercules helicopter} /  it is durable enough I can't even break it with my terrible skills. \\ \hline
     \scriptsize Bought for our \textbf{Granddaughter(she is 3)} for \textbf{Christmas}.  She just loves the write on wipe off A,B,C's and 1,2,3's.  The \textbf{art projects} that were included and quality of the items for the project, TERRIFIC!  Would recommend for all 3 year olds. & \scriptsize This is perfect for a rainy day \textbf{Christmas Vacation} / \textbf{my three yr old LOVES crafts} / Filled with all the supplies to make 16 high quality crafts \\ \hline
    \end{tabular}
    \label{tab:rel_review}
\end{table}

\vspace{-0.5em}
\section{Conclusion}
\vspace{-0.5em}
We propose a new model ZARM for the review based rating prediction task. In our model, the interaction module based on relevance matching function with zero-attention network is utilized to learn user dynamic representation in more flexible way. And the auxiliary loss plugged into the training object make the relevance matching function better trained. Experiments on the four Amazon benchmark datasets show our model can outperform the state-of-art models based on the siamese network and interaction-based network.
By conducting case studies, we take a deeper look into the behavior of our interaction module, and investigate the several statistical and semantic characteristics of the relevant reviews for users to targets extracted by the interaction module.
\vspace{-4pt}
\section*{Acknowledgements}
\vspace{-6pt}
This work was supported in part by the School of Computing, University of Utah and in part by NSF IIS-2007398. Any opinions, findings and conclusions or recommendations expressed in this material are those of the authors and do not necessarily reflect those of the sponsor. 

\newpage

\bibliographystyle{splncs04}
\bibliography{sigproc}

\end{document}